\documentclass[10pt,journal,twoside,final]{IEEEtran}

\usepackage{diagbox}
\usepackage{rotating}
\usepackage{graphicx,subfig}
\usepackage{cite}
\usepackage{enumitem}
\usepackage{color}
\usepackage{bigstrut}
\usepackage{multirow}

\usepackage{amsmath, amsthm, amssymb}
\usepackage{url}
\usepackage{array}
\usepackage{booktabs}
\newcolumntype{L}[1]{>{\raggedright\let\newline\\\arraybackslash\hspace{0pt}}m{#1}}
\newcolumntype{C}[1]{>{\centering\let\newline\\\arraybackslash\hspace{0pt}}m{#1}}
\newcolumntype{R}[1]{>{\raggedleft\let\newline\\\arraybackslash\hspace{0pt}}m{#1}}
\usepackage{amssymb}

\DeclareMathOperator*{\argmin}{argmin}

\newcommand{\etal}{\textit{et al. }}

\begin{document}

\title{Hiding Secrets in the CSI Quotient: A Robust Wi-Fi CSI Steganography System}

\author{
    Jiamu~Guo,
    Hailang~Jia,
    Guanxiong~Shen,~\IEEEmembership{Member,~IEEE},
	Junqing~Zhang,~\IEEEmembership{Senior Member,~IEEE},
    Linning~Peng,~\IEEEmembership{Member,~IEEE}, and
    Liquan~Chen,~\IEEEmembership{Senior Member,~IEEE}
	
	\thanks{Manuscript received xxx; revised xxx; accepted xxx. Date of publication xxx; date of current version xxx. 
    The review of this paper was coordinated by xxx. 
    \textit{(Corresponding author: Guanxiong Shen.)}}
    \thanks{J.~Guo, H.~Jia, G.~Shen, and L.~Chen are with the School of Cyber Science and Engineering, Southeast University, Nanjing, China. (email: jiamuguo@seu.edu.cn; hailang.jia@seu.edu.cn; gxshen@seu.edu.cn; lqchen@seu.edu.cn)}
	\thanks{J.~Zhang is with the School of Computer Science and Informatics, University of Liverpool, Liverpool, L69 3DR, United Kingdom. (email: junqing.zhang@liverpool.ac.uk)}
    \thanks{L.~Peng is with the School of Cyber Science and Engineering, Southeast University, Nanjing 210096, China, and also with Purple Mountain Laboratories, Nanjing 210096, China (email: pengln@seu.edu.cn)}

	\thanks{Color versions of one or more of the figures in this paper are available online at 
    xxx}
	\thanks{Digital Object Identifier xxx}
}

\maketitle

\begin{abstract}

Physical layer (PHY) steganography conceals secrets by making subtle modifications to transmitted radio waveforms, which can be applied to establish covert communication systems. Given the widespread deployment of Wi-Fi infrastructures, hiding secrets within Wi-Fi transmissions exhibits significant covertness and has attracted increasing research attention. 
Recent advances in Wi-Fi steganography have focused on embedding secrets within channel state information (CSI) by applying artificial finite impulse response (FIR) filters to outgoing signals. These methods can emulate natural wireless propagation effects, thereby evading detection by eavesdroppers. However, existing CSI-based approaches suffer from two critical limitations: vulnerability to environmental variations and limited steganographic capacity. This work presents a Wi-Fi steganography system that mitigates these constraints. Specifically, we introduce a CSI division mechanism to decouple artificial CSI components from natural wireless channel responses. In essence, secrets are embedded within the quotient of two consecutive CSI measurements. Furthermore, we propose an encoder-decoder neural network framework that automatically learns optimal strategies for FIR filter generation and secret recovery, substantially enhancing steganographic capacity. We implemented a prototype using commercial off-the-shelf hardware, including a software-defined radio (SDR) transmitter and two receiver platforms: ANTSDR and ESP32. Experimental evaluations demonstrate that the system achieves robust performance under dynamic environmental conditions while significantly improving steganographic capacity.

\end{abstract}


\section{Introduction}
 
Wireless steganography is a technique that conceals secret messages within ordinary communication links, which is regarded as an effective technique for achieving practical covert communication~\cite{jiang2024physical, bonati2021stealte, d2019hiding, chen2023covert}. Recent research has explored the application of steganographic methods to a variety of widely deployed wireless systems, such as 4G/5G cellular networks~\cite{bonati2021stealte}, ZigBee~\cite{liu2025no}, and LoRa~\cite{liu2023lophy, hou2022cloaklora}, to establish covert and hard-to-detect side channels for delivering sensitive information. Among these techniques, hiding secrets within Wi-Fi networks has attracted significant research interest~\cite{classen2015practical, chen2023covert, zou2016survey, rahbari2017exploiting, zhang2021adaptive}. As Wi-Fi has become the \textit{de facto} standard for wireless local area networks (WLANs), its global ubiquity enables the concealment of sensitive information within these pervasive Wi-Fi transmissions, thereby offering strong covertness and extensive coverage range.

Covert channels can be engineered at the physical layer (PHY). This approach, termed PHY steganography, refers to hiding information through deliberate, subtle manipulations of the radio waveform~\cite{hou2022cloaklora, liu2023lophy, liu2025no}. Recent literature has reported numerous proof-of-concept systems validating this paradigm. For instance, Wei~\etal employ a complex-valued neural network to generate covert signals that seamlessly mimic inherent hardware noise patterns~\cite{wei2023wise}. Similarly, the work in~\cite{d2019hiding} introduces a pseudo-noise asymmetric shift keying (PN-ASK) modulation scheme, which hides covert data by minimally adjusting the amplitude of primary in-phase/quadrature (I/Q) samples. These prototyping efforts underscore both the practicality and feasibility of implementing steganography directly at the wireless PHY. 

A promising direction in Wi-Fi PHY steganography leverages channel state information (CSI) as the covert carrier~\cite{schulz2018shadow, jiao2021openwifi}. Specifically, the transmitter applies an artificial finite impulse response (FIR) filter to the outgoing Wi-Fi signals, and the intended receiver analyzes the CSI to extract the hidden secrets~\cite{schulz2018shadow}.
The secret embedding process effectively superimposes a subtle, engineered channel signature onto the natural environmental channel.
A key advantage is that this artificial FIR filtering process does not interfere with the primary Wi-Fi transmissions. Conventional Wi-Fi receivers remain capable of correctly decoding the payload through standard channel equalization and demodulation procedures.
\textbf{The core reason for the undetectability of such CSI-based schemes lies in their capacity to mimic environmental wireless channels by applying an artificial FIR filter.}
As a result, it becomes highly challenging for a warden to distinguish whether the observed CSI is a measurement of the natural propagation environment or has been subtly manipulated via artificial FIR filtering.

Although prior prototype studies have experimentally validated the feasibility of embedding secrets within CSI, several significant challenges remain unsolved. 
First, the measured CSI represents a superposition of the embedded secrets and the characteristics of the surrounding wireless environment. As a result, any changes in the environment, such as relocation of the transceivers, the presence of moving objects, or variations in channel conditions, can significantly degrade the reliability and fidelity of the covert channel. For this reason, the authors in~\cite{jiao2021openwifi} argue that CSI-based steganography is only practical in relatively static environments. 
Second, existing Wi-Fi steganographic systems suffer from limited data hiding capacity. For example, the Shadow Wi-Fi system proposed in~\cite{schulz2018shadow} can only embed two secret bits per Wi-Fi frame, which constrains its applicability in scenarios requiring higher covert data throughput. These challenges motivate the development of a Wi-Fi steganography framework that is both robust to environmental variability and capable of supporting higher secret transmission throughput.

In this study, we designed an environmental-resilient Wi-Fi steganography system, which embeds secrets within CSI by applying an artificial FIR filter to the outgoing Wi-Fi waveforms. Specifically, we propose to embed secrets in the quotient of two consecutive CSI measurements. Given that the propagation channel remains relatively stable in low-mobility environments, this quotient allows for effective separation of the environmental and artificial components of CSI through a simple division operation. Furthermore, we employ a pair of neural networks to learn the FIR filter generation and secret recovery rules, which can enhance the steganographic capacity of the covert communication link. The main contributions of this work are summarized as follows.
\begin{itemize}
    \item We propose a practical Wi-Fi steganography system that hides secrets within CSI to mimic natural wireless propagation effects. Specifically, the transmitter Alice applies artificial FIR filters to two consecutive Wi-Fi signals, while Bob extracts secrets by analyzing the CSI measurements.
    \item The designed Wi-Fi steganography system is robust to environmental variations. The artificial CSI is separated from the environmental effects using a dedicated CSI divider, leveraging the slow-varying characteristic of the wireless channel in quasi-static environments. In other words, the secrets are embedded within the quotient of two consecutive CSIs.
    \item A pair of neural networks, i.e., encoder-decoder, is utilized to generate FIR coefficients and extract secrets from CSI measurements. This approach enables the neural networks to automatically learn an optimal secret embedding rule in simulation platforms, thus enhancing the steganographic capacity.
    \item We presented a prototype implementation using commodity devices. A software-defined radio (SDR) serves as the Wi-Fi transmitter, while the performance was evaluated on two distinct receiver platforms, including an SDR board, ANTSDR, and an Internet of things (IoT) development kit, ESP32. The experiment results demonstrated on these commercially available platforms validate the practicality of the proposed Wi-Fi steganography system.
\end{itemize}

\textbf{The reproducible code and experiment dataset will be made publicly available upon acceptance of this manuscript.}

The rest of the paper is organized as follows: Section~\ref{sec:system_overview} presents the system overview and high-level working principles. Section~\ref{sec:secret_embedding} details Alice's secret embedding operations. Section~\ref{sec:secret_extraction} elaborates Bob's secret extraction process. The neural network architecture and training scheme are given in Section~\ref{sec:neural_network_training}. Section~\ref{sec:experiment_evaluation} illustrates the experiment evaluation using commodity devices. The related work and conclusion are given in Section~\ref{sec:related_work} and Section~\ref{sec:conclusion}, respectively.

\section{System Overview}\label{sec:system_overview}

\begin{figure*}[!t]
  \centering
  \includegraphics[width = 6.8in]{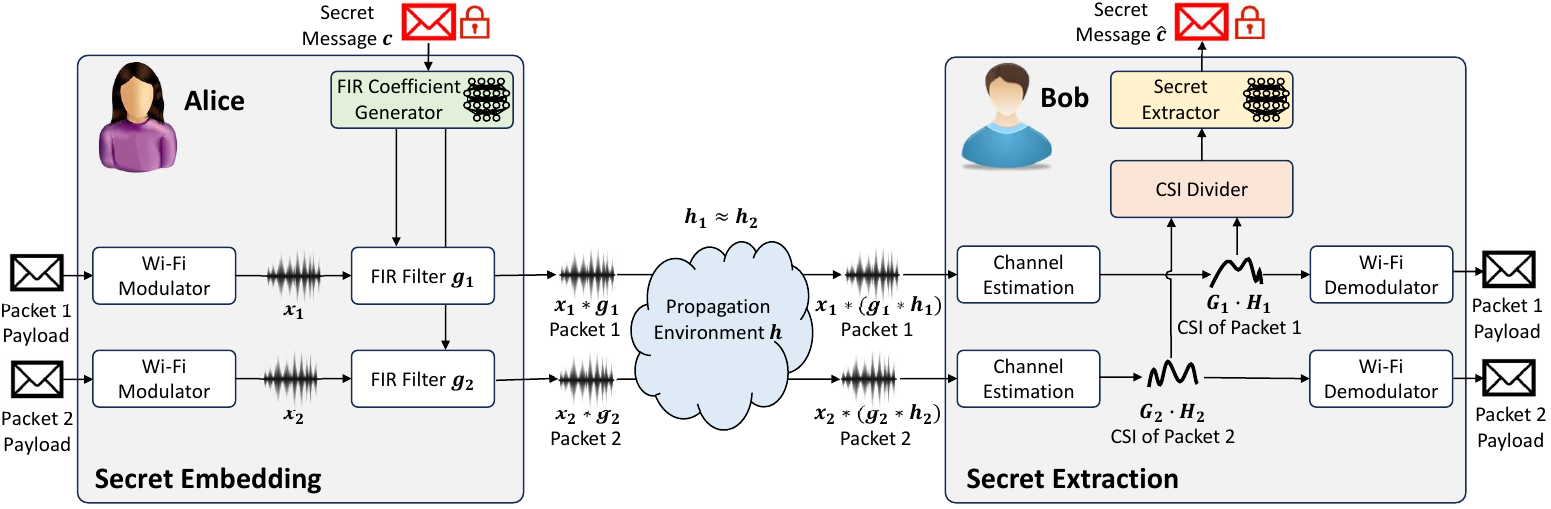}
  \caption{Overview of the Wi-Fi steganography system.}
  \label{fig:system_overview}
\end{figure*}

Fig.~\ref{fig:system_overview} presents the overall architecture of the proposed Wi-Fi steganography system. Two legitimate users, Alice and Bob, first establish a standard Wi-Fi link for normal data transmission. Building on this existing connection, Alice transmits additional secret messages through a covert PHY channel, while ensuring that the standard Wi-Fi payload reception remains unaffected.

\textbf{The key idea is to embed secrets into the CSI by introducing controlled artificial multipath effects.} Specifically, Alice applies two intentionally crafted FIR filters to the PHY waveforms of two consecutive Wi-Fi packets. 
The following subsections describe Alice’s embedding process, Bob’s extraction process, and the design and training of the neural networks that enable this system.

\subsection{Overview of Alice's Operations}

At the transmitter, Alice embeds secret messages by introducing subtle distortions into two consecutive Wi-Fi packets. Given a secret bit sequence, a neural network, termed the \emph{FIR coefficient generator}, maps the secret message to a pair of FIR filter coefficient vectors, denoted as $\mathbf{g}_1$ and $\mathbf{g}_2$. 

Subsequently, Alice applies $\mathbf{g}_1$ to the PHY waveform of the first Wi-Fi packet, and $\mathbf{g}_2$ to that of the immediately following packet via FIR filtering modules. This operation introduces artificial multipath effects that produce secret-bearing patterns in the resulting CSI measurements. Meanwhile, a standard Wi-Fi receiver can still perform channel estimation and equalization to correctly decode the original payload, leaving the primary Wi-Fi link unaffected.

Optionally, forward error correction (FEC) is applied to the secret bits prior to embedding to improve robustness against channel noise and CSI estimation errors.


\subsection{Overview of Bob's Operations}

At the receiver side, Bob captures the two consecutive Wi-Fi transmissions. Each received signal undergoes natural wireless propagation as well as the artificial filtering imposed by Alice. By performing standard channel estimation, Bob obtains the CSI of the two packets, denoted as $\mathbf{\hat{C}}_1$ and $\mathbf{\hat{C}}_2$. Each CSI thus reflects a superposition of the environmental channel response and Alice's artificial filter.

To disentangle the secret-bearing artificial components from the uncontrollable environmental channel, Bob exploits the temporal stability of wireless channels in quasi-static environments. Specifically, a \emph{CSI divider} computes the quotient between $\mathbf{\hat{C}}_1$ and $\mathbf{\hat{C}}_2$, effectively canceling the common environmental propagation effects while preserving the relative features introduced by $\mathbf{g}_1$ and $\mathbf{g}_2$. As a result, the obtained CSI quotient primarily carries information about the embedded secrets.

Finally, Bob feeds this quotient into another neural network, termed the \emph{secret extractor}, which maps the CSI quotient back to the original secret bits. Throughout this entire process, Bob can still decode the original Wi-Fi payload via standard Wi-Fi demodulation, ensuring that the covert communication does not interfere with normal data transmission and remains fully compatible with off-the-shelf Wi-Fi devices.

\subsection{Neural Networks Training and Fine-Tuning}

The FIR coefficient generator and secret extractor constitute the core learning components of the proposed Wi-Fi steganography system. The generator maps secret messages to paired FIR filters for waveform embedding, while the extractor reconstructs the embedded secrets from the CSI quotient at the receiver.
Both networks are jointly trained on a simulated platform, as described in Section~\ref{sec:training_process}. This end-to-end optimization allows the model to automatically learn coordinated embedding and decoding strategies that maximize secret recovery reliability.

After training, the generator is deployed at Alice and the extractor at Bob. An external observer without access to the trained extractor parameters cannot retrieve the embedded secrets, thereby providing an additional security layer beyond PHY-level concealment. Although the training is conducted in simulation, practical hardware inevitably introduces non-idealities, such as sampling offsets and RF impairments. To alleviate potential simulation-to-reality mismatch, we optionally fine-tune the extractor using a small amount of measurement data collected on the target receiver platform, enabling better compensation for hardware distortions, as detailed in Section~\ref{sec:fine-tuning}. Notably, fine-tuning is not mandatory: experiments show that even without fine-tuning, the simulation-trained model achieves strong real-world performance, demonstrating the robustness of the proposed end-to-end framework.

\section{Secret Embedding by Alice}\label{sec:secret_embedding}

This section details Alice's operations in the Wi-Fi steganography system. Alice's goal is to embed a secret message into two consecutive Wi-Fi packets by mimicking natural wireless channel effects. This makes the covert transmission undetectable to potential eavesdroppers. The secret embedding process consists of two main steps: generating FIR filter coefficients using a neural network, and applying these filters to Wi-Fi waveforms.

\subsection{Principle of CSI Secret Embedding}\label{sec:prop_model}

This subsection provides a concise signal model to clarify \emph{how the secret is embedded into the PHY waveform} and \emph{why the embedding remains covert}. 
Our key observation is that multipath propagation can be modeled as an FIR filter in baseband. By deliberately applying artificial FIR filters at the transmitter, we effectively create synthetic multipath components that are indistinguishable from natural propagation characteristics.

\subsubsection{Wireless Channel Model}

We consider the discrete-time complex baseband representation of the Wi-Fi PHY waveform, where each packet consists of a sequence of I/Q samples. 
Let $\mathbf{x}$ denote the transmitted baseband samples and $\mathbf{y}$ denote the received samples. 
The multipath propagation environment can be modeled by a (possibly time-varying) channel impulse response (CIR) $\mathbf{h}$, given by
\begin{equation}\label{equ:wireless_channel_discrete}
    \mathbf{y} = \mathbf{h} * \mathbf{x} + \mathbf{n},
\end{equation}
where $(*)$ denotes discrete convolution and $\mathbf{n}$ is additive noise. 
\eqref{equ:wireless_channel_discrete} is the standard discrete-time baseband model used by Wi-Fi receivers for synchronization and channel estimation.

\subsubsection{Secret Embedding as an Artificial Multipath}

In our system, Alice embeds the secret by applying an artificial FIR filter $\mathbf{g}$ to the baseband waveform before transmission. 
For two consecutive packets, this operation is
\begin{equation}\label{equ:tx_filtering_recap}
  \left\{
    \begin{aligned}
      \mathbf{z}_1 = \mathbf{g}_1 * \mathbf{x}_1,
      \\
      \mathbf{z}_2 = \mathbf{g}_2 * \mathbf{x}_2.
    \end{aligned}
  \right.
\end{equation}
where $\mathbf{x}_1$ and $\mathbf{x}_2$ represent the complex baseband I/Q sample sequences of the first and second Wi-Fi packets, respectively, and $\mathbf{g}_1$ and $\mathbf{g}_2$ are the corresponding artificial FIR filters applied by Alice.
After propagation through the environment, Bob receives

\begin{equation}\label{equ:received_signals_discrete}
  \left\{
    \begin{aligned}
      \mathbf{y}_1 = \mathbf{h}_1 * \mathbf{z}_1 + \mathbf{n}_1 = (\mathbf{h}_1 * \mathbf{g}_1) * \mathbf{x}_1 + \mathbf{n}_1,
      \\
      \mathbf{y}_2 = \mathbf{h}_2 * \mathbf{z}_2 + \mathbf{n}_2 = (\mathbf{h}_2 * \mathbf{g}_2) * \mathbf{x}_2 + \mathbf{n}_2.
    \end{aligned}
  \right.
\end{equation}

\eqref{equ:received_signals_discrete} shows that the receiver observes a CSI resulting from the cascade of the environmental channel $\mathbf{h}_i$ and the artificial filter $\mathbf{g}_i$, i.e., $\mathbf{h}_i * \mathbf{g}_i$. 
Consequently, $\mathbf{g}_i$ can be viewed as additional multipath components superimposed on the original CIR $\mathbf{h}_i$. 
Under this condition, the presence of artificial filtering is naturally concealed within the environmental propagation effects.

\subsection{FIR Coefficient Generator}\label{sec:coefficient_generator}

Having established the signal model, we now describe how Alice generates the artificial FIR filters that carry the secret information. 
A neural network named FIR coefficient generator is employed to map the secret message $\mathbf{c}$ to two sets of FIR filter coefficients, denoted as $\mathbf{g}_1$ and $\mathbf{g}_2$, which are mathematically given as
\begin{equation}
    \{\mathbf{g}_1, \mathbf{g}_2\} = \mathcal{G}(\mathbf{c}; \theta_{\mathcal{G}}), 
\end{equation}
where $\mathcal{G}(\cdot; \theta_{\mathcal{G}})$ represents the coefficient generator, with $\theta_{\mathcal{G}}$ being its parameters. The detailed architecture of the coefficient generator is elaborated in Section~\ref{sec:nn_design}. The vectors $\mathbf{g}_1$ and $\mathbf{g}_2$ are two sets of FIR filter coefficients produced by the neural network, each comprising $L$ taps, and are defined as
\begin{equation}
  \left\{
    \begin{aligned}
      \mathbf{g}_1 = [g_{1}^1, g_{1}^2, \cdots, g_{1}^{L}],
      \\
      \mathbf{g}_2 = [g_{2}^1, g_{2}^2, \cdots, g_{2}^{L}],
    \end{aligned}
  \right.
\end{equation}
where $g_{1}^i$ and $g_{2}^i$ are the $i$-th tap of $\mathbf{g}_1$ and $\mathbf{g}_2$, respectively. Note that each tap is a complex number, making both $\mathbf{g}_1$ and $\mathbf{g}_2$ complex vectors. Each secret message $\mathbf{c}$ is mapped to a unique pair of $\mathbf{g}_1$ and $\mathbf{g}_2$.

\subsection{FIR Filter: Secret Embedding through Filtering}\label{sec:filtering}

After generating the FIR filter coefficients $\mathbf{g}_1$ and $\mathbf{g}_2$, they are applied to two consecutive Wi-Fi packets by the FIR filter module, respectively, given as
\begin{equation}\label{equ:tx_filtering}
  \left\{
    \begin{aligned}
      \mathbf{z}_1 = \mathbf{g}_1 * \mathbf{x}_1,
      \\
      \mathbf{z}_2 = \mathbf{g}_2 * \mathbf{x}_2,
    \end{aligned}
  \right.
\end{equation}
where $\mathbf{x}_1$ and $\mathbf{x}_2$ represent the PHY I/Q samples of two successive Wi-Fi packets output from the Wi-Fi modulator. The outputs, $\mathbf{z}_1$ and $\mathbf{z}_2$, are the filtered signals that are emitted into the air.

\section{Secret Extraction by Bob}\label{sec:secret_extraction}

This section details Bob's operations to extract the embedded secret messages from the received Wi-Fi signals. Bob faces a significant challenge: the CSI he measures contains both the secret-bearing artificial channel effects introduced by Alice and the natural environmental channel effects. To successfully recover the secret, Bob must distangle these two components. This section describes how Bob accomplishes this through three main steps: channel estimation, CSI division, and secret extraction using a neural network.

\subsection{Channel Estimation: Extracting CSI from Received Signals}

Wi-Fi systems employ Orthogonal Frequency Division Multiplexing (OFDM), which divides the available bandwidth into multiple orthogonal subcarriers. This multi-carrier modulation scheme provides robustness against frequency-selective fading and enables efficient spectrum utilization. The Wi-Fi receiver measures the wireless channel in the frequency domain across these subcarriers, commonly referred to as CSI or channel frequency response (CFR). 

Channel estimation is a standard module in the Wi-Fi receiver chain, which compensates for the distortion introduced by the wireless channel during signal propagation. For a 20~MHz Wi-Fi system, the estimated CSI is a vector composed of 52 complex values, corresponding to 52 subcarriers. The CSI can be obtained by applying a fast Fourier transform (FFT) to the channel impulse response, given as
\begin{equation}
    \mathbf{H} = \text{FFT}(\mathbf{h}),
\end{equation}
where $\mathbf{H}$ represents the estimated CSI in the frequency domain, and $\text{FFT}(\cdot)$ denotes the fast Fourier transform operator\footnote{In practice, CSI extraction entails a sequence of signal-processing steps, including FFT, FFT shifting, and subcarrier selection (e.g., retaining 52 of the 64 subcarriers in 20~MHz Wi-Fi). For simplicity, we omit these implementation details.}. In essence, CSI provides a frequency-domain measurement of the wireless channel characteristics, which are determined by the propagation environment and exhibit time-varying behavior due to the mobility of transceivers and dynamic changes in surrounding objects.

In the proposed Wi-Fi steganography system, the estimated CSI reflects a composite channel response that incorporates both the artificial filtering effect intentionally introduced by Alice and the underlying environmental channel characteristics. This can be mathematically represented as
\begin{equation}
  \left\{
    \begin{aligned}
      \mathbf{\hat{C}}_1 = \text{FFT}(\mathbf{g}_1 * \mathbf{h}_1) + \mathbf{W} = \mathbf{G}_1 \odot \mathbf{H}_1 + \mathbf{W},
      \\
      \mathbf{\hat{C}}_2 = \text{FFT}(\mathbf{g}_2 * \mathbf{h}_2) + \mathbf{W} = \mathbf{G}_2 \odot \mathbf{H}_2 + \mathbf{W},
    \end{aligned}
  \right.
\end{equation}
where $\mathbf{\hat{C}}_1$ and $\mathbf{\hat{C}}_2$ are the CSIs estimated from the received Wi-Fi signals $\mathbf{y}_1$ and $\mathbf{y}_2$, respectively. $\mathbf{G}_1$, $\mathbf{G}_2$, $\mathbf{H}_1$, and $\mathbf{H}_2$ are the frequency-domain representations of $\mathbf{g}_1$, $\mathbf{g}_2$, $\mathbf{h}_1$, and $\mathbf{h}_2$, respectively. The operator $\odot$ denotes element-wise multiplication. $\mathbf{W}$ represents the CSI estimation error resulting from noise.

Notably, the secret message $\mathbf{c}$ is hidden within $\{\mathbf{G}_1, \mathbf{G}_2\}$, while $\mathbf{H}_1$ and $\mathbf{H}_2$ represent environmental effects. The challenge for Bob is to extract $\{\mathbf{G}_1, \mathbf{G}_2\}$ from the composite measurements $\{\mathbf{\hat{C}}_1, \mathbf{\hat{C}}_2\}$ without knowing $\mathbf{H}_1$ and $\mathbf{H}_2$.

\subsection{CSI Divider: Mitigating Environmental Effects}\label{sec:csi_divider}
A critical challenge in CSI-based steganography is that the measured CSI contains both the embedded secret (through $\mathbf{G}_1$ and $\mathbf{G}_2$) and the environmental channel effects ($\mathbf{H}_1$ and $\mathbf{H}_2$). To reliably extract the secret, we need to remove the environmental component. Our key insight is to exploit the temporal stability of wireless channels in quasi-static environments.

\subsubsection{CSI Characteristics in Quasi-Static Environments}
A quasi-static environment refers to a wireless communication scenario where the channel characteristics change slowly compared to the transmission timescale. In such an environment, the wireless channel is assumed to remain approximately constant over a short period of time, typically on the order of microseconds to milliseconds. This assumption is particularly valid in scenarios with limited or slow-moving scatterers around the transmitter and receiver, such as indoor settings characterized by stationary objects and regular human activity. Under these conditions, the time-domain channel impulse response $\mathbf{h}$ exhibits slow temporal variations, and correspondingly, the frequency-domain CSI $\mathbf{H}$ can be considered relatively stable across consecutive Wi-Fi packets.

\begin{figure}[!t]
  \centering
  \includegraphics[width = 3.3in]{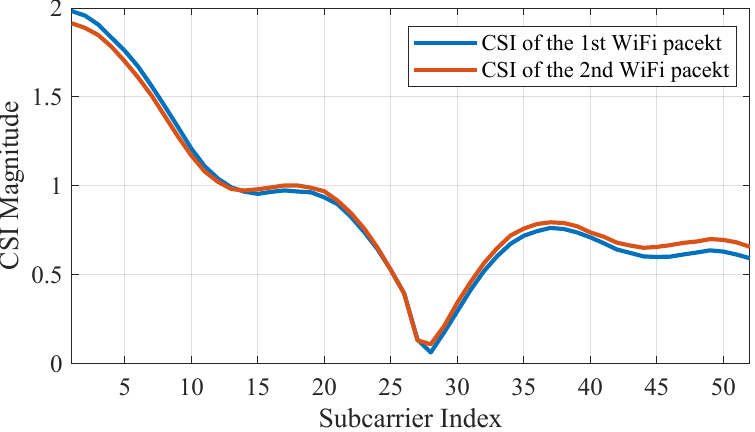}
  \caption{The CSI results in a quasi-static simulation environment. TGn Model-C is used with an environmental speed of 20~m/s. The interval between Wi-Fi packets is 158~$\mu$s.}
  \label{fig:csi_quasi}
\end{figure}

To illustrate the characteristics of CSI in quasi-static environments, we present MATLAB simulations based on the TGn channel model-C\footnote{The scenario for TGn model-C is indoor residential or small office.} provided in the MATLAB WLAN Toolbox~\cite{matlab_wlan_toolbox}. The environmental mobility speed is set to 20~m/s, and a noise level of 30~dB is simulated. To evaluate the temporal stability of the channel, we simulate two consecutive Wi-Fi transmissions following the IEEE 802.11n standard, with the resulting CSI estimations presented in Fig.~\ref{fig:csi_quasi}. The simulation results clearly demonstrate the quasi-static nature of the wireless environment. As observed in the figure, compared to the CSI of the first packet, the channel remains nearly unchanged after 158 $\mu$s, i.e., 108~$\mu$s packet airtime and 50~$\mu$s interval. This confirms the slow-varying property of the wireless channel in the absence of rapidly moving objects, laying the foundation for our CSI division approach.

\subsubsection{CSI Division Operation}
Upon receiving the two consecutive Wi-Fi signals and estimating their respective CSIs, receiver Bob performs element-wise division of the two CSI measurements to mitigate environmental channel effects, which is expressed as
\begin{equation}\label{equ:csi_divider}
      \mathbf{\hat{C}}_{div} = \frac{\mathbf{\hat{C}}_1}{\mathbf{\hat{C}}_2} = \frac{\mathbf{G}_1 \odot \mathbf{H}_1}{\mathbf{G}_2 \odot \mathbf{H}_2},
\end{equation}
where $\mathbf{\hat{C}}_{div}$ represents the division result, and the division is performed element-wise across all subcarriers. 
According to the discussion presented above, it is reasonable to assume that the wireless channel is relatively stable over the short time interval between two consecutive packets. This implies that $\mathbf{H}_1$ and $\mathbf{H}_2$ are approximately identical, given as
\begin{equation}
  \mathbf{H}_1 \approx \mathbf{H}_2.
\end{equation}
Consequently,~\eqref{equ:csi_divider} can be approximated as
\begin{equation}
  \mathbf{\hat{C}}_{div} \approx \frac{\mathbf{G}_1}{\mathbf{G}_2}.
\end{equation}
Through this division operation, the environmental channel effects $\mathbf{H}_1$ and $\mathbf{H}_2$ are effectively canceled out, and the resulting quotient $\mathbf{\hat{C}}_{div}$ primarily contains the information related to the secret message $\mathbf{c}$ embedded within the artificial FIR filter coefficients. This is the key innovation that enables robust secret extraction across different environments.

\subsection{Secret Extractor: Recovering the Secret Messages}

After mitigating the environmental effects through the division of the two CSI measurements, receiver Bob extracts the secret message $\mathbf{c}$ from the division result $\mathbf{\hat{C}}_{div}$ using a pre-trained neural network termed the secret extractor. The secret extraction process can be mathematically formulated as
\begin{equation}
  \mathbf{\hat{c}} = \mathcal{E}(| \mathbf{\hat{C}}_{div} |; \theta_{\mathcal{E}}),
\end{equation}
where $\mathcal{E}(\cdot; \theta_{\mathcal{E}})$ denotes the secret extractor, and $\theta_{\mathcal{E}}$ is the corresponding parameter set. $\mathbf{\hat{c}}$ is the recovered secret message. It should be noted that only the magnitude of $\mathbf{\hat{C}}_{div}$, denoted as $| \mathbf{\hat{C}}_{div} |$, is fed into the neural network, as phase information is inherently sensitive to noise.

\section{Neural Network Training and Fine-Tuning}\label{sec:nn_design}\label{sec:neural_network_training}

The neural networks, i.e., the FIR coefficient generator $\mathcal{G}(\cdot; \theta_{\mathcal{G}})$ and the secret extractor $\mathcal{E}(\cdot; \theta_{\mathcal{E}})$, constitute the core components of the proposed Wi-Fi steganography system. The architectural design and training methodologies are detailed in this section.

\subsection{Neural Network Architectures}

\begin{figure}[!t]
  \centering
  \includegraphics[width = 3.4in]{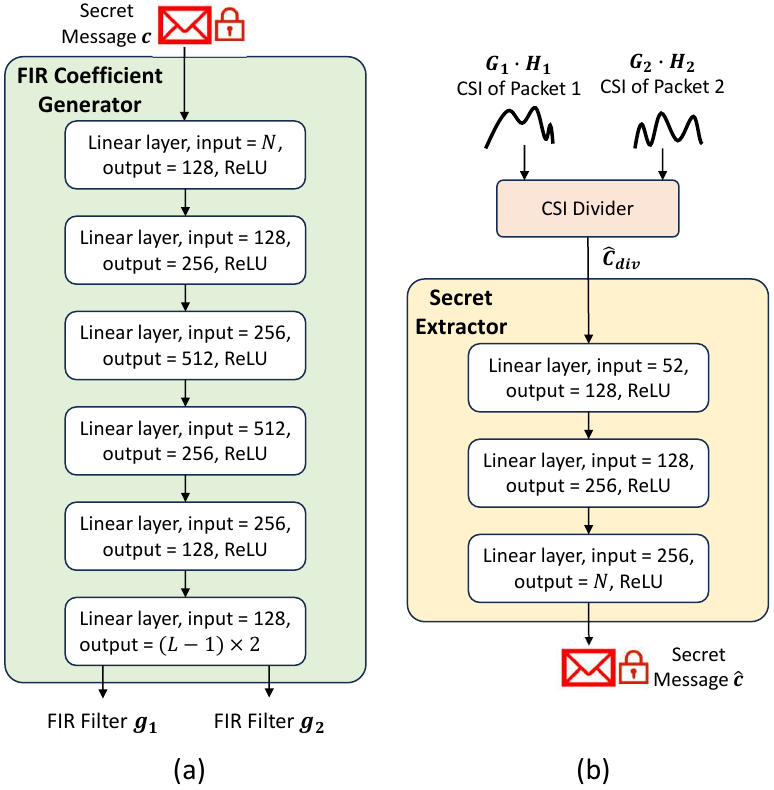}
  \caption{The architecture of the neural networks. (a) FIR coefficient generator. (b) Secret extractor.}
  \label{fig:neural_network_architecture}
\end{figure}

The neural network architecture is depicted in Fig.~\ref{fig:neural_network_architecture}. Both the FIR coefficient generator $\mathcal{G}(\cdot; \theta_{\mathcal{G}})$ and the secret extractor $\mathcal{E}(\cdot; \theta_{\mathcal{E}})$ are implemented as multilayer perceptrons (MLPs), a well-established deep learning architecture. While more sophisticated neural network architectures are available, their increased implementation complexity must be carefully evaluated against the practical constraints of commodity wireless hardware platforms.

The FIR coefficient generator $\mathcal{G}(\cdot; \theta_{\mathcal{G}})$ consists of six linear layers, accepting the secret message as input and producing two sets of FIR filter coefficients, $\mathbf{g}_1$ and $\mathbf{g}_2$. Since $\mathbf{g}_1$ and $\mathbf{g}_2$ are complex-valued vectors, while neural networks operate on real-valued data, $\mathcal{G}(\cdot; \theta_{\mathcal{G}})$ outputs the concatenated real and imaginary components, which are subsequently transformed into complex-valued coefficients. Furthermore, a Tanh activation function is applied to the output of $\mathcal{G}(\cdot; \theta_{\mathcal{G}})$ to constrain the coefficient values to the range $[-1, 1]$, while other layers are activated by the ReLU function.

The secret extractor $\mathcal{E}(\cdot; \theta_{\mathcal{E}})$ comprises three linear layers, accepting the magnitude of the division result $|\mathbf{\hat{C}}_{div}|$ as input and producing the recovered secret message. The output of each layer is activated by the ReLU function.

\subsection{End-to-End Neural Network Training}\label{sec:training_process}
\begin{figure}[!t]
  \centering
  \includegraphics[width = 2.6in]{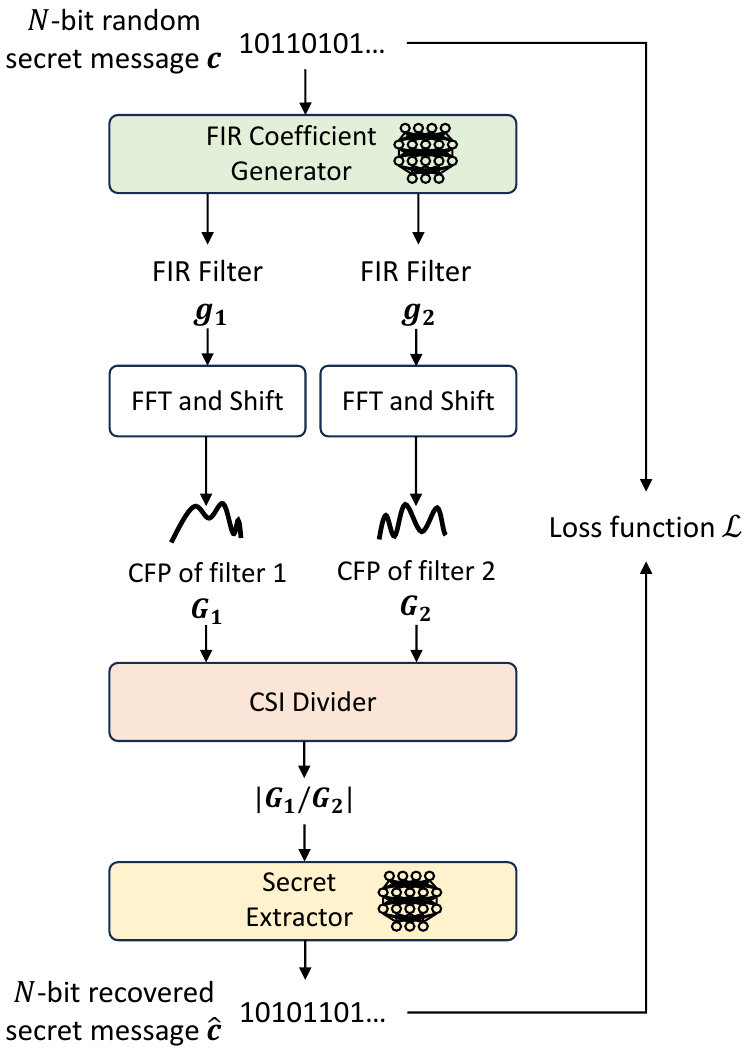}
  \caption{Training methodology of the neural networks, emulating a complete communication process.}
  \label{fig:training_scheme}
\end{figure}

The training of the neural networks, illustrated in Fig.~\ref{fig:training_scheme}, is conducted in an end-to-end manner. This approach ensures that the entire training pipeline replicates the secret embedding and extraction procedure, with a loss function designed to guide the iterative weight updates.

The training procedure begins with the random generation of a batch of $N$ binary numbers, which represent the secret message $\mathbf{c}$. These bits are subsequently fed as input to the FIR coefficient generator $\mathcal{G}(\cdot; \theta_{\mathcal{G}})$, which generates two sets of filter coefficients, $\mathbf{g}_1$ and $\mathbf{g}_2$.

In the next stage, we apply a 64-point FFT to both $\mathbf{g}_1$ and $\mathbf{g}_2$, converting the FIR coefficients to frequency-domain CSI. It is worth noting that an FFT shift operation is also applied, and the output is subsequently truncated to 52 subcarriers to align with standard wireless communication protocols.
The pair of CSIs is then fed into the CSI divider for processing. Subsequently, the secret extractor $\mathcal{E}(\cdot; \theta_{\mathcal{E}})$ processes the output from the CSI divider and reconstructs the secret message $\mathbf{\hat{c}}$. To enhance the system's robustness, artificial noise is introduced to $\mathbf{ | \hat{C}}_{div} | $ during training.

A recovery loss $\mathcal{L}$ is defined to guide the weight updating process, which quantifies the discrepancy between $\mathbf{\hat{c}}$ and ground-truth $\mathbf{c}$, mathematically expressed as
\begin{equation}\label{equ:loss}
  \mathcal{L} = \frac{1}{B}\sum_{i=1}^{B}\frac{\| \mathbf{\hat{c}}_i - \mathbf{c}_i \|_2}{N},
\end{equation}
where $B$ denotes the batch size and $\| \cdot \|_2$ returns the L2 norm. $\mathbf{\hat{c}}_i$ and $\mathbf{c}_i$ are the $i$-th recovered and ground-truth secret message within a batch, respectively.
The objective of neural network training is to determine the optimal parameters $\theta_{\mathcal{G}}$ and $\theta_{\mathcal{E}}$ that minimize the recovery loss $\mathcal{L}$, given as
\begin{equation}
  \{ \theta_{\mathcal{G}}, \theta_{\mathcal{E}} \}= \mathop{\argmin}_{\{ \theta_{\mathcal{G}}, \theta_{\mathcal{E}} \}} \sum_{\{ \mathbf{\hat{c}}, \mathbf{c} \} \in \mathcal{S}} \mathcal{L}(\{ \mathbf{\hat{c}}, \mathbf{c} \}; \{ \theta_{\mathcal{G}}, \theta_{\mathcal{E}} \}),
\end{equation}
where $\mathcal{S}$ denotes the sample space of the $N$-bit recovered secret message $\mathbf{\hat{c}}$ and ground-truth $\mathbf{c}$. In practice, the training process is conducted using gradient-based optimization methods, which iteratively update the network parameters to minimize the recovery loss.

\subsection{Neural Network Fine-Tuning}\label{sec:fine-tuning}

\begin{figure}[!t]
  \centering
  \includegraphics[width = 3.2in]{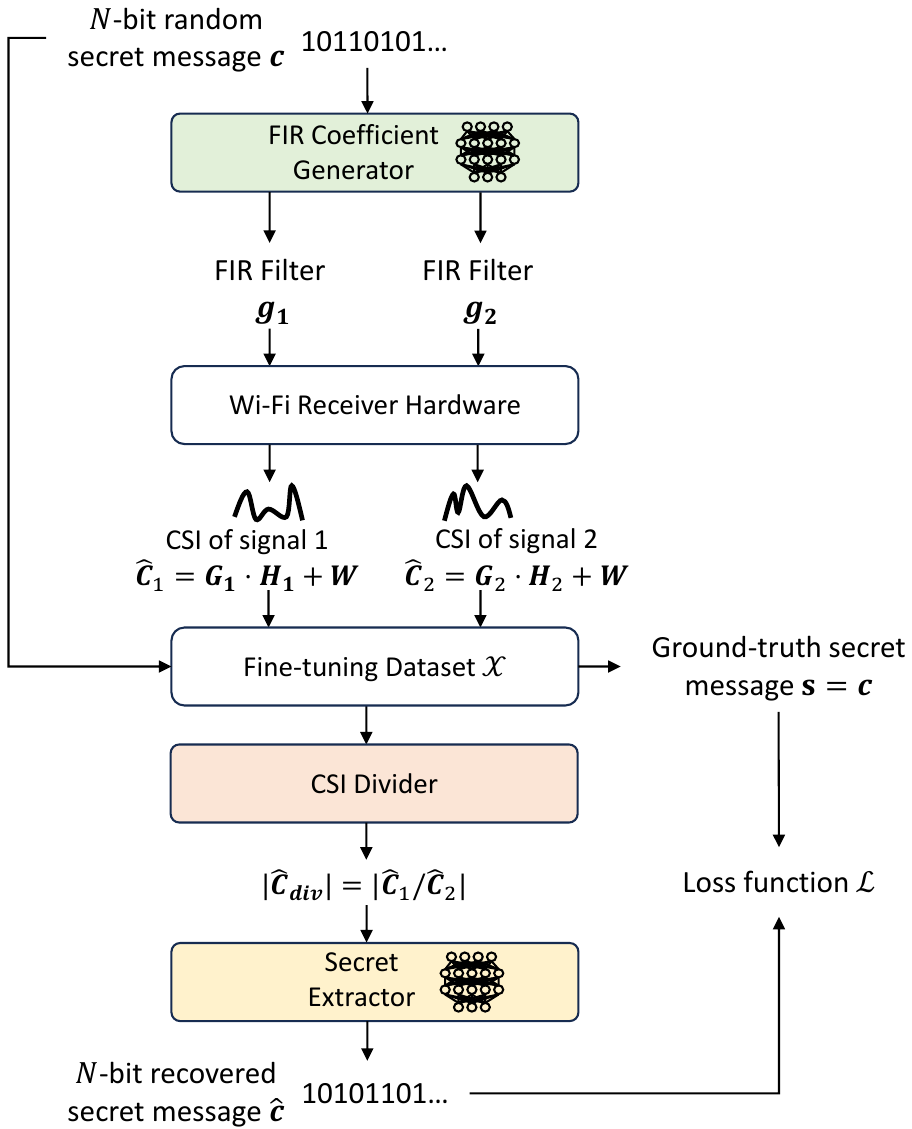}
  \caption{Fine-tuning scheme of the neural networks, using actual collected datasets.}
  \label{fig:fine-tuning_scheme}
\end{figure}

The training procedure described above relies entirely on simulated data without incorporating real-world measurements. However, neural network performance inevitably degrades on actual hardware platforms due to hardware imperfections. To address this issue, we propose a fine-tuning scheme to enhance the performance of the secret extractor.

As illustrated in Fig.~\ref{fig:fine-tuning_scheme}, we first deploy the pre-trained neural network models on actual hardware platforms and then collect a fine-tuning dataset $\mathcal{X}$, given as 
\begin{equation}
  \mathcal{X} = \{ ((\mathbf{\hat{C}_1}, \mathbf{\hat{C}_2})_r, \mathbf{s}_r ) \}_{r=1}^{R},
\end{equation}
where $(\mathbf{\hat{C}_1}$, $\mathbf{\hat{C}_2})_r$ denotes the CSIs extracted from the $r$-th pair of collected Wi-Fi packets. $\mathbf{s}_r$ represents the corresponding ground-truth secret codeword. $R$ denotes the number of samples in the fine-tuning dataset. After collecting a sufficient number of fine-tuning samples, we compute $\mathbf{\hat{C}}_{div}$ and feed it to the secret extractor to calibrate the parameter set $\theta_{\mathcal{E}}$ using the loss function $\mathcal{L}$ defined in (\ref{equ:loss}).

Note that fine-tuning is an optional step for enhanced performance. The experimental results in Section~\ref{sec:experiment_evaluation} demonstrate that neural networks trained solely with simulated data already exhibit excellent performance on real hardware platforms.

\section{Experiment Evaluation}\label{sec:experiment_evaluation}
\begin{figure}[!t]
  \centering
  \includegraphics[width = 2.7in]{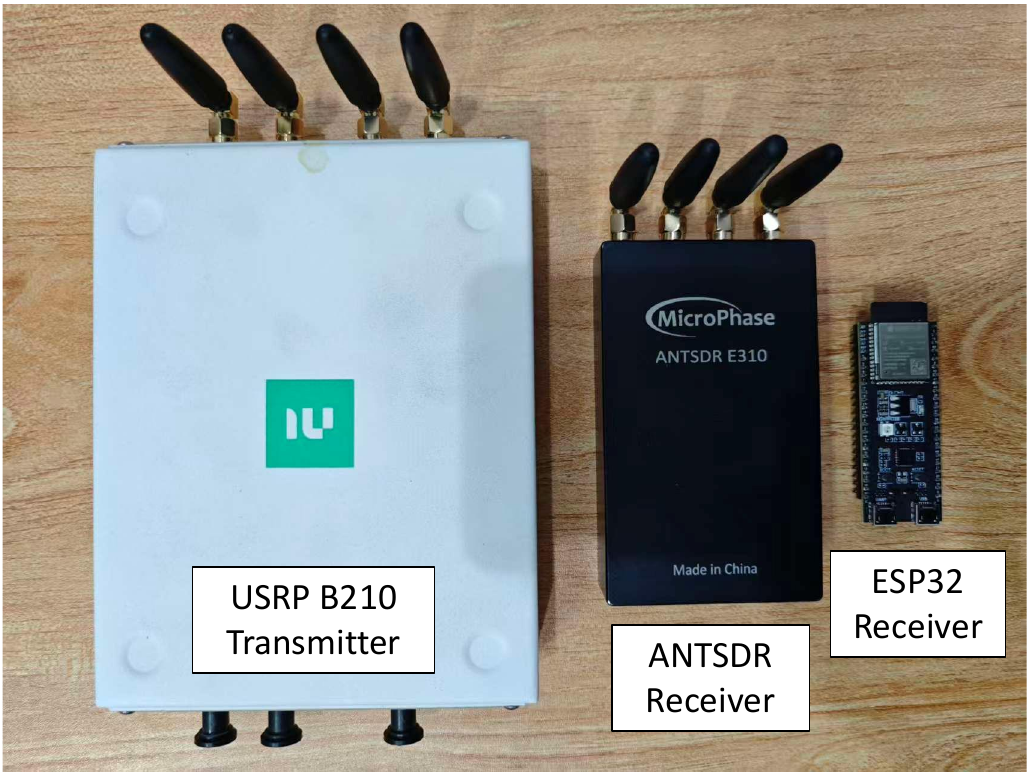}
  \caption{Experimental devices. A USRP B210 transmitter, an ANTSDR receiver, and an ESP32 receiver.}
  \label{fig:experiment_device}
\end{figure}

As shown in Fig.~\ref{fig:experiment_device}, we created a Wi-Fi steganography prototype using a USRP B210 transmitter and two distinct receivers, i.e., ANTSDR and ESP32. The evaluation results are detailed in this section.

\subsection{Experiment Setup}

\subsubsection{Wi-Fi Transmitter Setup}

We developed a Wi-Fi transmitter utilizing a USRP B210 SDR, enabling the direct application of custom-designed FIR filters to the PHY waveform. The USRP SDR was connected to a laptop running the Wi-Fi PHY and secret embedding programs. Specifically, neural network-generated FIR filters were sequentially applied to Wi-Fi baseband signals, and the resulting filtered waveforms were subsequently up-converted and transmitted over-the-air via the USRP B210.

The Wi-Fi PHY implementation utilizes the MATLAB WLAN Toolbox to continuously generate IEEE 802.11n beacon frames\footnote{The proposed steganography scheme is applicable to any Wi-Fi frame type, with beacon frames used as a case example.}. The system operates at a center frequency of 2.462 GHz, i.e., Wi-Fi Channel 11, with a bandwidth of 20 MHz. For identification purposes, the transmitter's MAC address is configured as 00:12:34:56:78:9B, and the packet interval is set to 158 $\mu$s.

\subsubsection{Wi-Fi Receiver Setup}

To implement the proposed Wi-Fi steganography system, Wi-Fi receivers must provide access to CSI. This functionality is supported by various readily available open-source tools and commercial off-the-shelf Wi-Fi devices. To comprehensively evaluate the proposed system, we conducted experiments using two distinct hardware platforms: ANTSDR and ESP32, which are described below.
\begin{itemize}
  \item \textbf{ANTSDR Receiver Setup}: 
  We employed an ANTSDR E200 platform to capture the Wi-Fi signals, which was equipped with an AD9361 RF transceiver chip. We installed the open-source OpenWiFi project on an SDR board~\cite{jiao2020openwifi, openwifi_github}, making it function as a standard Wi-Fi network interface card (NIC) while providing access to PHY data such as CSI. The SDR was configured to operate in monitor mode, enabling passive capture of Wi-Fi packets. The ANTSDR was connected to a host laptop via USB, through which both CSI measurements and timestamps were streamed for subsequent analysis. During CSI and timestamp acquisition, we concurrently executed a separate tcpdump instance on the ANTSDR to capture and parse packet-level information from received Wi-Fi frames. The CSI data and packet information were subsequently aligned based on their respective timestamps.
  
  \item \textbf{ESP32 Receiver Setup}: The ESP32 is a low-cost microcontroller unit (MCU) with integrated Wi-Fi functionality, commonly employed in IoT applications. The ESP32 provides official application programming interfaces (APIs) to access CSI measurements~\cite{esp32_github}. In our implementation, we utilized an ESP32-S3 development kit configured in monitor mode to collect CSI data from incoming Wi-Fi packets. The captured CSI measurements were subsequently transmitted to a host laptop via serial connection for processing and analysis.
  
\end{itemize}

\subsubsection{Neural Network Configuration}

The neural networks, namely the FIR coefficient generator and the secret extractor, were deployed on the host laptops at the transmitter and receiver, respectively. The implementation was conducted using the PyTorch framework. During the training process, the RMSProp optimizer was utilized with a batch size $B$ of 64 and an initial learning rate of 0.0003. The learning rate was adaptively reduced by a factor of 0.1 when the validation loss failed to decrease over 10 consecutive epochs. Additionally, training was terminated using early stopping if no reduction in validation loss was observed for 20 consecutive epochs.
Since the neural networks are trained in a physically controlled environment, only the transmitter and receiver possess the weights of the FIR coefficient generator and the secret extractor, respectively.

\subsubsection{Evaluation Metrics}

The BER is calculated directly from the Hamming distance between the transmitted and extracted secret message $\mathbf{c}$ and $\mathbf{\hat{c}}$, given as
\begin{equation}
    BER = \frac{\mathcal{D}(\mathbf{c}, \mathbf{\hat{c}})}{N},
\end{equation}
where $\mathcal{D}(\cdot, \cdot)$ denotes the Hamming distance between two binary vectors, and $N$ is the length of the secret message. The BER serves as a direct indicator of the neural network's performance.

\subsection{Evaluation of Steganographic Capacity}

This section evaluates the maximum amount of hidden data that can be reliably transmitted by the Wi-Fi steganography system, i.e., steganographic capacity. In the experiments, a USRP B210 transmitter continuously transmitted Wi-Fi beacon frames with random secret messages embedded. Two different platforms were used for signal reception: ANTSDR and ESP32.

\begin{figure}[!t]
  \centering
  \includegraphics[width = 3.3in]{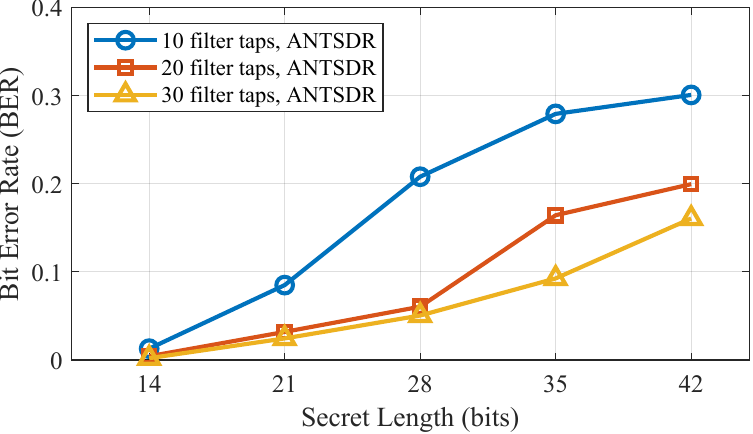}
  \caption{Evaluation of steganographic capacity on ANTSDR receiver.}
  \label{fig:capacity_openwifi}
\end{figure}

\textbf{ANTSDR Receiver:} The steganographic capacity performance on the ANTSDR platform is illustrated in Fig.~\ref{fig:capacity_openwifi}. As demonstrated by the experimental results, the BER exhibits an increasing trend with the growth of secret message length $N$. Specifically, when utilizing 30 coefficients, the BER increases from 0.0021 to 0.16 as the secret length extends from 14 bits to 42 bits. This observed trend is reasonable, as the embedding and extraction of longer secret messages inherently present greater challenges. The experimental results also demonstrate that increasing the number of FIR taps $L$ yields enhanced steganographic capacity. For instance, when the secret message length is fixed at 21 bits, the BER decreases significantly from 0.085 to 0.0245 as the number of taps increases from 10 to 30. This performance enhancement can be attributed to the expanded hiding capacity afforded by additional FIR taps, which provides the neural network with greater degrees of freedom for secret embedding.

\begin{figure}[!t]
  \centering
  \includegraphics[width = 3.3in]{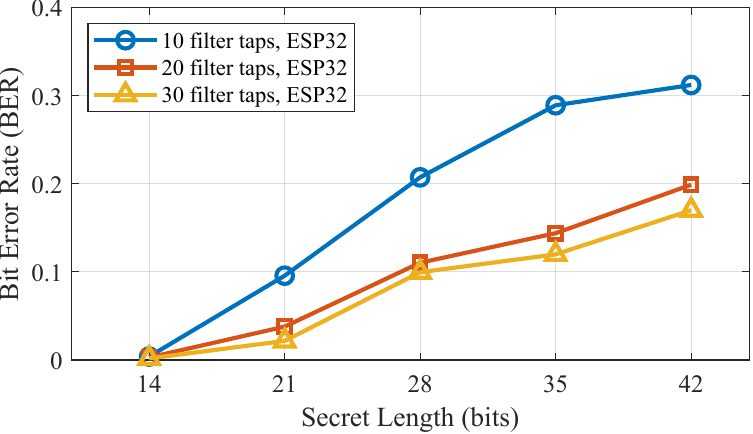}
  \caption{Evaluation of steganographic capacity on ESP32 receiver.}
  \label{fig:capacity_esp32}
\end{figure}

\textbf{ESP32 Receiver:} The evaluation results obtained on the ESP32 platform are illustrated in Fig.~\ref{fig:capacity_esp32}. The results exhibit trends consistent with those observed on the ANTSDR platform. Specifically, the BERs increase with longer secret message lengths and decrease as the number of FIR filter taps increases, thereby validating the generalizability of the proposed approach across different hardware platforms. In the best-performing configuration, where 30 FIR taps are used to embed a 14-bit secret, the system achieves remarkable performance metrics with a BER of 0.0018. These results represent near-perfect error-free communication, demonstrating the feasibility of the proposed Wi-Fi steganography method on low-cost embedded platforms.

The experimental results across both platforms confirm that the designed system can achieve a BER of 0.16 when embedding 42 bits into the Wi-Fi CSI.
Although transmission errors are present, the application of FEC techniques can effectively mitigate transmission errors. For instance, the Bose–Chaudhuri–Hocquenghem (BCH) code can theoretically detect and correct errors of up to nearly 25\%~\cite{al2017enhancing, zhang2016efficient}. The hiding performance demonstrates a clear positive correlation with the number of FIR taps generated by the neural network, as more taps provide greater flexibility for secret embedding.

\subsection{Evaluation of Environmental Robustness}\label{sec:evaluation_environment}

One of the key challenges addressed by this work is ensuring system robustness against environmental variations. To address this challenge, we employed a CSI divider that effectively eliminates ambient environmental CSI components while preserving the artificial ones, as detailed in Section~\ref{sec:csi_divider}.

This section presents experimental validation of the system's robustness against environmental variations. We conducted comprehensive experiments across both indoor and outdoor environments using two distinct Wi-Fi receivers: ANTSDR and ESP32. The number of FIR taps $L$ is set to 30, and the length of secret messages $N$ is set to 14. The detailed results are presented in the following subsections.

\begin{figure}[!t]
	\centering
    
    \subfloat[]{\includegraphics[width=2.7in]{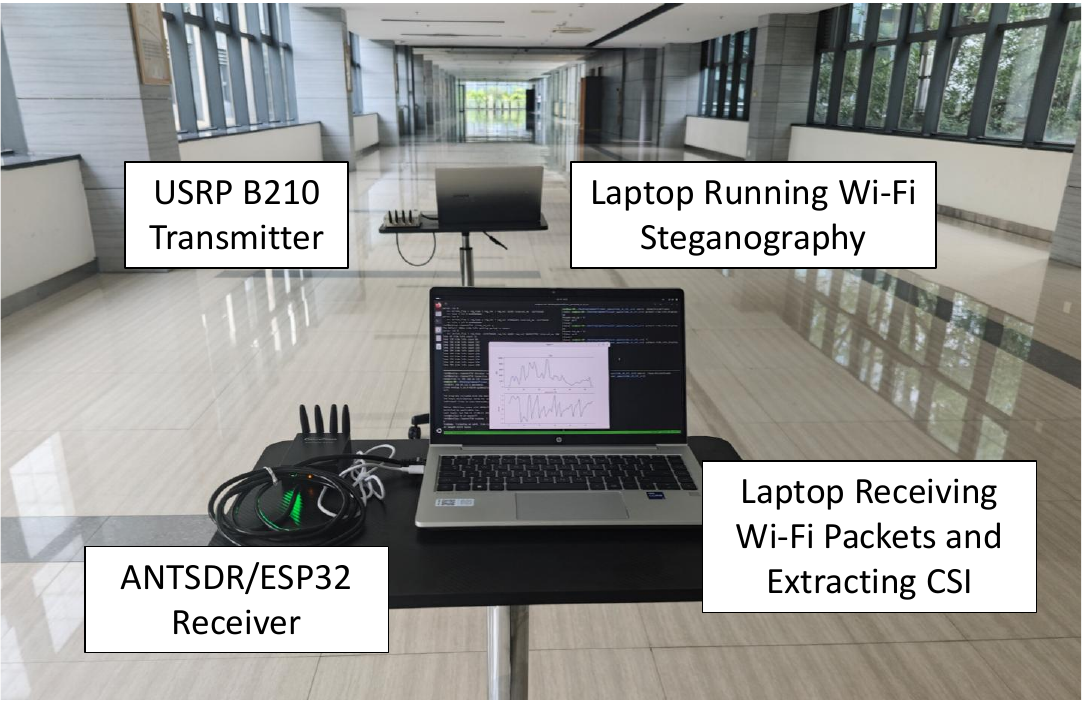}
        \label{}}
        
    \subfloat[]{\includegraphics[width=2.7in]{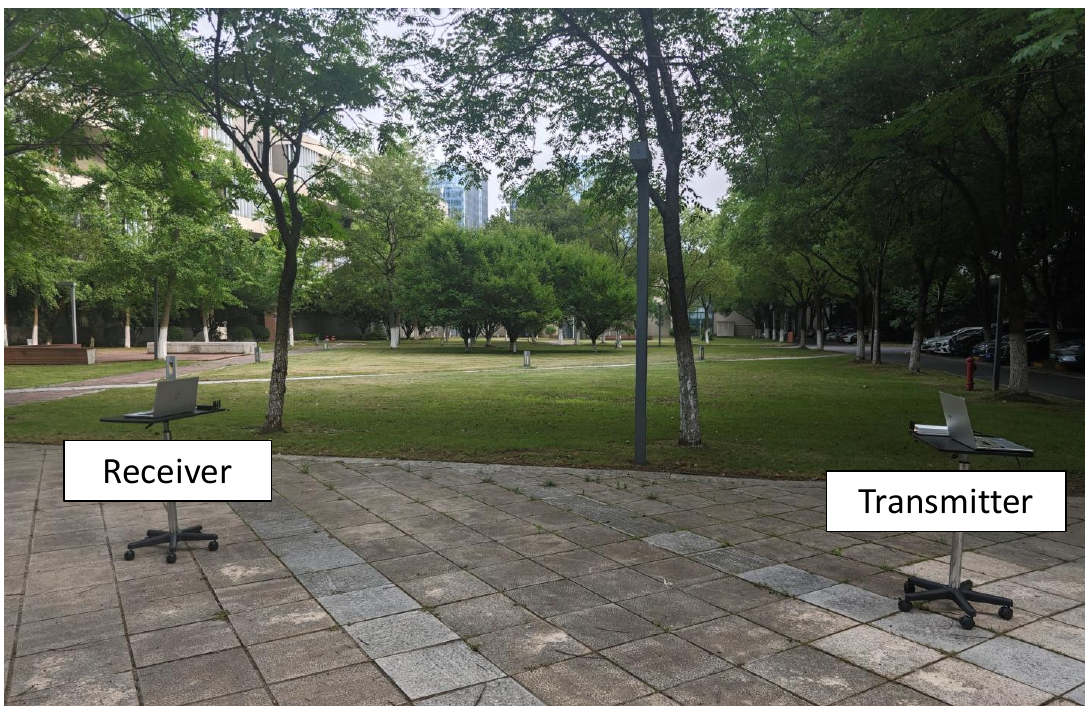}
        \label{}}

	\caption{The experiment environments. (a) Indoor scenario inside an office building. (b) Outdoor scenario inside a park.}
	\label{fig:experiment_environment}
\end{figure}

\begin{figure}[!t]
	\centering
    
    \subfloat[]{\includegraphics[width=1.7in]{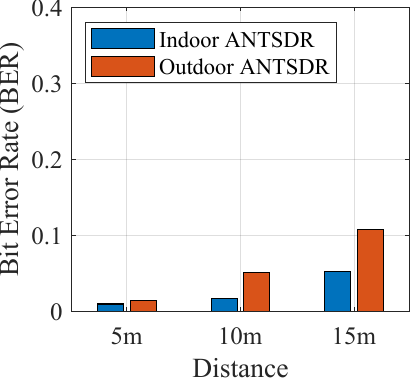}
        \label{}}
    \subfloat[]{\includegraphics[width=1.7in]{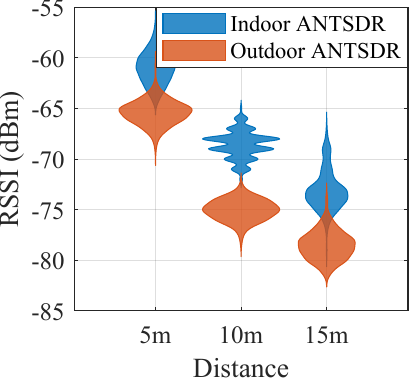}
        \label{}}

	\caption{Evaluation of environmental robustness on ANTSDR receiver. (a) Bit error rate during secret message recovery. (b) Received signal strength indicator (RSSI).}
	\label{fig:environment_openwifi}
\end{figure}

\textbf{ANTSDR Receiver:} 
We first evaluated the system performance in an indoor environment, as shown in Fig.~\ref{fig:experiment_environment}(a). The SDR transmitter and ANTSDR receiver were placed in a corridor with separation distances of 5m, 10m, and 15m. The experimental results are presented in Fig.~\ref{fig:environment_openwifi}. The covert Wi-Fi system demonstrates satisfactory performance across all tested distances. The BER exhibits an upward trend as transmission distance increases. Specifically, when the distance increases from 5m to 15m, the BER rises from 0.0096 to 0.0518. This degradation is primarily attributed to the deterioration of SNR conditions with increasing distance, as the average RSSI declines from -60.99 dBm to -73.1 dBm in Fig.~\ref{fig:environment_openwifi}(b). We conduct similar experiments in an outdoor environment, as shown in Fig.~\ref{fig:experiment_environment}(b), with results presented as orange bars in Fig.~\ref{fig:environment_openwifi}. The system performance degrades significantly compared to the indoor setting, primarily due to worse SNR conditions in the outdoor environment. In the most severe case, at a transmission distance of 15~m, the BER reaches 0.1075.

\begin{figure}[!t]
	\centering
    
    \subfloat[]{\includegraphics[width=1.7in]{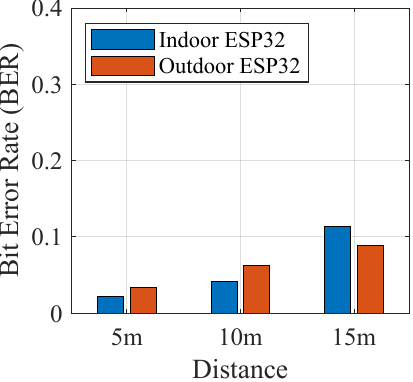}
        \label{}}    
    \subfloat[]{\includegraphics[width=1.7in]{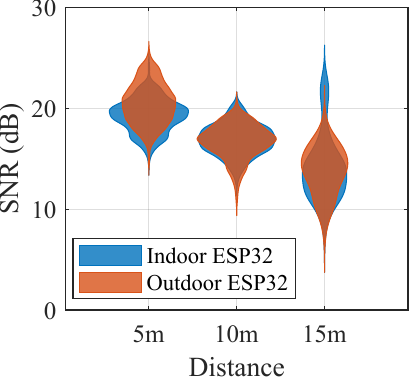}
        \label{}}

	\caption{Evaluation of environmental robustness on ESP32 receiver. (a) Bit error rate during secret message recovery. (b) Estimated SNR values.}
	\label{fig:environment_esp32}
\end{figure}
    
\textbf{ESP32 Receiver:} 
We replicated the experiments using the ESP32 device, with results shown in Fig.~\ref{fig:environment_esp32}. The findings closely align with those obtained using the ANTSDR receiver. System performance demonstrates a strong correlation with signal quality, as reflected by the SNR. Under adverse conditions at a 15-meter distance, the BER reaches 0.1143 and 0.0893 for indoor and outdoor environments, respectively.

The experimental results demonstrate the robustness of the proposed Wi-Fi steganography system under varying environmental conditions. The embedded covert data can be reliably extracted across diverse ambient scenarios. Moreover, the results reveal a negative correlation between the covert link quality and the transceiver distance, which is expected, as increased propagation distance generally results in signal attenuation and a corresponding reduction in the SNR at the receiver.

\subsection{Effect of Fine-Tuning}

The neural networks employed in Section~\ref{sec:evaluation_environment} are trained using simulated data, while real-world hardware imperfections also affect secret extraction performance. In this section, we deploy the fine-tuning scheme described in Section~\ref{sec:fine-tuning} and evaluate its effectiveness.

Specifically, we collect real-world datasets using two distinct receivers: an ANTSDR and an ESP32, both operating with a 14-bit secret message length as configured in Section~\ref{sec:evaluation_environment}. For each receiver, we gather 500 pairs of Wi-Fi beacon frames in an indoor environment at a 1-meter transmission distance. These datasets are used to fine-tune the original simulation-trained model, employing the Adam optimizer with a batch size of 32 and an initial learning rate of 0.00003. The resulting two fine-tuned secret extractors are then evaluated across all real-world datasets described in Section~\ref{sec:evaluation_environment}, which include varied distances and both indoor and outdoor settings.

\begin{table}[!t]
  \centering
  \caption{The fine-tuning results for the ANTSDR and ESP32 receiver when the secret message length is 14 bits.}
    \begin{tabular}{clcc}
    \toprule
    \multirow{2}[2]{*}{Dataset} & \multirow{2}[2]{*}{Model} & \multicolumn{2}{c}{BER} \\  \cmidrule{3-4}   &     & ANTSDR & ESP32 \\ \midrule
    \multirow{2}[2]{*}{Indoor 5 m} & Simulation & 0.0096  & 0.0218 \\
          & Fine-tuned & 0.0054  & 0.0157 \\
    \midrule
    \multirow{2}[1]{*}{Indoor 10 m} & Simulation & 0.0168  & 0.0411 \\
          & Fine-tuned & 0.0104  & 0.0329  \\
    \midrule
    \multirow{2}[1]{*}{Indoor 15 m} & Simulation & 0.0518  & 0.1143 \\
          & Fine-tuned & 0.0450  & 0.1039 \\
    \midrule
    \multirow{2}[2]{*}{Outdoor 5 m} & Simulation & 0.0139  & 0.0343  \\
          & Fine-tuned & 0.0107  & 0.0286  \\
    \midrule
    \multirow{2}[1]{*}{Outdoor 10 m} & Simulation & 0.0507  & 0.0625  \\
          & Fine-tuned & 0.0464  & 0.0546  \\
    \midrule
    \multirow{2}[1]{*}{Outdoor 15 m} & Simulation & 0.1075  & 0.0893  \\
          & Fine-tuned & 0.0961  & 0.0821  \\
    \bottomrule
    \end{tabular}%
  \label{tab:14bits_fine-tune}%
\end{table}%

The BERs for both the original simulation-trained model and the fine-tuned models are summarized in Table~\ref{tab:14bits_fine-tune}. Across all test conditions, the fine-tuned models consistently outperform their simulation-only counterparts. For instance, in the outdoor 15~m scenario, fine-tuning reduces the BER from 0.1075 to 0.0961 for the ANTSDR and from 0.0893 to 0.0821 for the ESP32. This improvement persists regardless of environmental variations or receiver type, demonstrating that fine-tuning effectively compensates for hardware-specific distortions that are absent in synthetic training data. The results thus confirm the practical value of the fine-tuning strategy in bridging the sim-to-real gap and enhancing robustness in real-world hardware platforms.

\subsection{Impact on Primary Wi-Fi Links}

A critical requirement for wireless steganography systems is that they must not interfere with primary Wi-Fi communications. In this section, we validate this essential property by using a commercial off-the-shelf smartphone to receive beacon frames transmitted by the SDR-based transmitter. 

\begin{figure}[!t]
    \centering
    \includegraphics[width=3.4in]{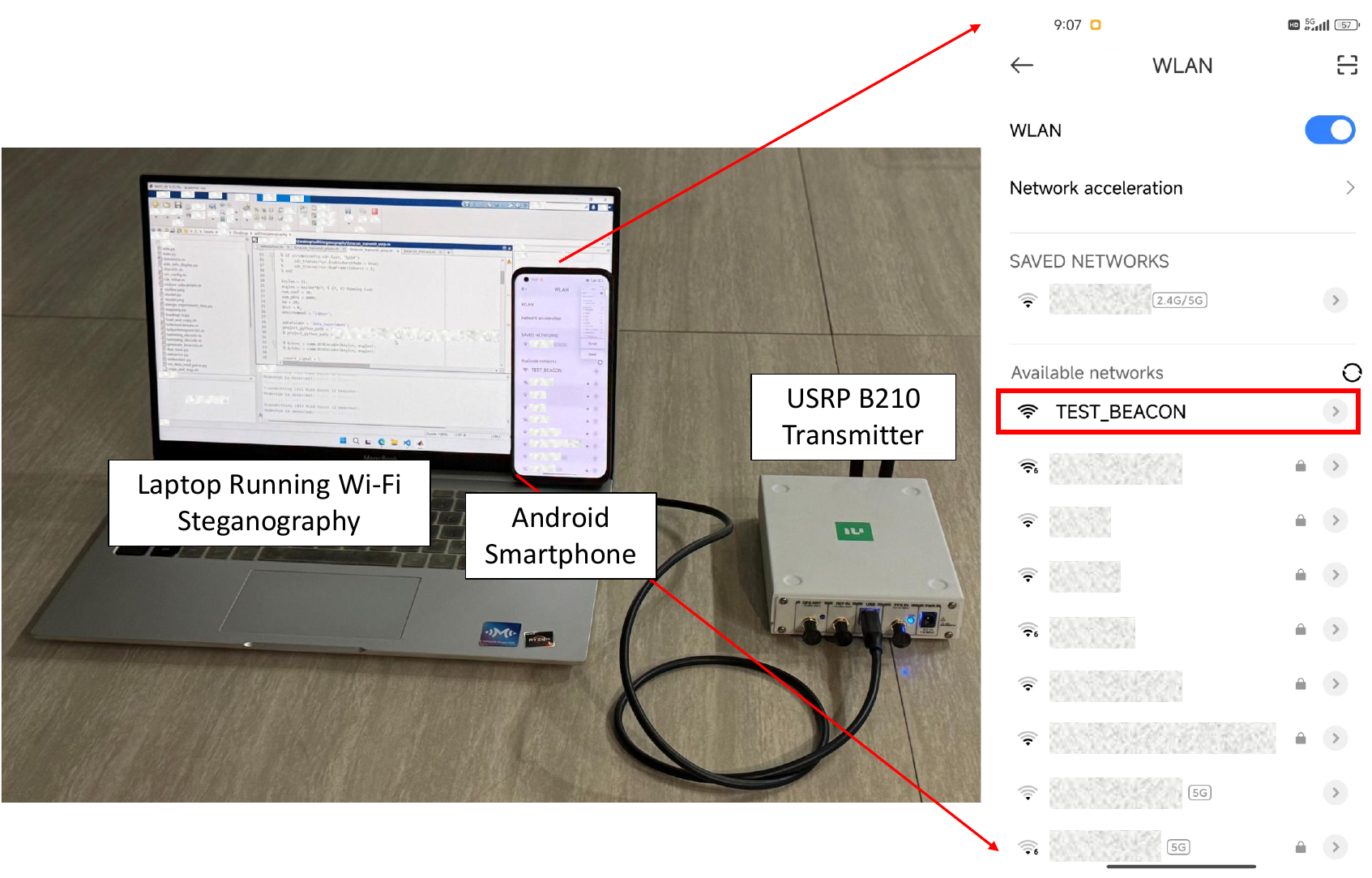}
    \caption{An SDR-based transmitter broadcasts Wi-Fi beacon frames with random secret messages embedded. While an Android smartphone successfully discovers the Wi-Fi AP named `TEST\_BEACON' in the network discovery interface.}
    \label{fig:smartphone}
\end{figure}

As shown in Fig.~\ref{fig:smartphone}, a USRP B210 device was configured to continuously broadcast Wi-Fi beacon frames on channel 11. These beacons were assigned the SSID `TEST\_BEACON' and carried a randomly generated 21-bit secret embedded within each pair of consecutive frames. The smartphone successfully detected the transmitted signal, with the configured SSID clearly displayed in the list of available networks. These results confirm that primary Wi-Fi functionality remains unaffected by the covert data embedding.

In the proposed Wi-Fi steganography system, Alice and Bob establish a covert communication channel overlaid on a standard Wi-Fi link. While regular users can still access and monitor the primary communication channel, only Bob is able to extract the embedded secret messages. This experiment confirms the system's capability to enable invisible and non-intrusive covert communication.

\section{Related Work}\label{sec:related_work}

Covert communication refers to the technique of concealing the existence of wireless links, which is particularly valuable for applications where undetectability is critical~\cite{chen2023covert, jiang2024physical}. Existing covert communication approaches can be broadly categorized into two main categories.
The first category employs directional transmission to spatially focus signals to intended receivers. This approach utilizes advanced technologies such as reconfigurable intelligent surfaces~\cite{lu2020intelligent, wang2021intelligent,zhou2021intelligent,wang2022intelligent,wang2024intelligent,wu2023intelligent} or multiple antenna arrays~\cite{shmuel2021multi,du2022performance,zheng2019multi} to implement spatial beamforming, thereby precisely directing signals to the locations of legitimate receivers while minimizing signal leakage to unintended areas. However, these methods often rely on idealized assumptions, such as perfect channel state information, and involve considerable implementation complexity, which limits their validation primarily to theoretical analysis and simulations.
The second category focuses on intentionally interfering with potential eavesdroppers to protect the covert communication channel. This can be accomplished through various techniques, including the emission of carefully designed artificial noise by the transmitter, or the deployment of friendly jammers that selectively interfere with potential eavesdroppers without disrupting the legitimate communication link~\cite{zheng2021wireless,zhang2021covert, huang2021jamming, jin2025covert}. While these studies provide rigorous theoretical analysis, practical prototype implementations are rarely demonstrated due to implementation complexity.

Wireless PHY steganography represents another promising approach to implement effective covert communication, achieved by embedding additional secret information within PHY wireless waveforms~\cite{sankhe2019impairment, d2019hiding, bonati2021stealte, wei2023wise, dutta2012secret}. This technique is fundamentally based on information hiding principles, wherein covert data is hidden within ordinary media~\cite{petitcolas2002information, katzenbeisser2016information, zhu2018hidden}. Recent studies have investigated its application across various wireless protocols, including cellular networks~\cite{d2019hiding, bonati2021stealte}, Wi-Fi~\cite{schulz2018shadow, jiao2021openwifi, classen2015practical}, LoRa~\cite{liu2023lophy, hou2022cloaklora}, and ZigBee~\cite{liu2025no} systems. Among these protocols, Wi-Fi stands out as particularly attractive due to its ubiquity in modern society. The widespread presence of Wi-Fi infrastructures provides an ideal cover medium, as secrets can be embedded within these ubiquitous Wi-Fi signals, rendering improved covertness and coverage range.

Deep learning has emerged as a powerful tool for steganography. Compared to hand-crafted rules designed by domain experts, deep learning models can automatically discover optimal embedding and extraction strategies, achieving significantly higher steganographic capacity~\cite{lu2021large, duan2020new}. By carefully designing the training process to incorporate realistic distortions such as JPEG compression, these models can learn effective embedding rules that are robust to common real-world perturbations. Moreover, generative adversarial networks (GANs) and other advanced generative models further enhance covertness by producing steganographic media that closely resemble natural, unmodified signals~\cite{zhang2019steganogan, yang2019embedding}.
While deep learning has significantly advanced steganographic techniques in other media such as image~\cite{d2019hiding, jing2021hinet, baluja2019hiding, bai2024information}, text~\cite{abdelnabi2021adversarial, peng2023cross, yang2018rnn}, and audio~\cite{wu2020audio}, its application to wireless systems remains largely unexplored.

\section{Conclusion}\label{sec:conclusion}

This study presented a Wi-Fi steganography system with improved environmental robustness and capacity. Two critical limitations in CSI-based steganography were considered, namely environmental vulnerability and limited steganographic capacity. By introducing a CSI divider mechanism and leveraging encoder-decoder neural networks for optimized filter design and secret extraction, our approach achieves robustness against dynamic propagation environments while significantly enhancing secret data throughput. We implemented a Wi-Fi steganography prototype on commodity hardware platforms, including a USRP transmitter and two distinct receivers, i.e., ANTSDR and ESP32. Experimental evaluations were conducted in both indoor and outdoor scenarios with various transceiver distances. The results demonstrate that even under varying environmental conditions and across different receiver platforms, the BER remains below 0.12 when embedding 14-bit secret messages. The prototype-driven evaluation demonstrates practical feasibility under real-world conditions.

\bibliographystyle{IEEEtran}
\bibliography{IEEEabrv,mybibfile}

\end{document}